\documentclass[a4paper,fleqn]{cas-dc}

\usepackage[numbers, sort&compress]{natbib}
\usepackage{physics}
\setcitestyle{square}

\definecolor{green_cust}{HTML}{00FF00}

\definecolor{blue_cust}{HTML}{0066ff}

\definecolor{gold_cust}{HTML}{FFBF00}

\definecolor{purple_cust}{HTML}{800080}
\newcommand{\purple}[1]{\textcolor{purple_cust}{#1}}

\definecolor{pink_cust}{HTML}{DC267F}

\definecolor{perfect_green}{HTML}{4FBF26}

\newcommand{\LEOF}[1]{{ \textbf{Comment from Leo}: }\{\purple{#1}\}}

\def\tsc#1{\csdef{#1}{\textsc{\lowercase{#1}}\xspace}}
\tsc{WGM}
\tsc{QE}
\tsc{EP}
\tsc{PMS}
\tsc{BEC}
\tsc{DE}

\newcommand{\red}[1]{\textcolor{crimson}{#1}}
\definecolor{crimson}{HTML}{DC143C}
\newcommand{\TODO}[1]{{ \textbf{ToDo}: }\{\red{#1}\}}
\begin{document}
\let\WriteBookmarks\relax
\def\floatpagepagefraction{1}
\def\textpagefraction{.001}

\shorttitle{On the use of the Axelrod formula in Astrophysical Modelling}

\shortauthors{Leo P. Mulholland et~al.}

\title [mode = title]{On the use of the Axelrod formula for thermal electron collisions in Astrophysical Modelling}                      
\tnotemark[1]

\tnotetext[1]{This document is the results of the research
   project funded by the European Union (ERC, HEAVYMETAL, 101071865). Views and opinions expressed are however those of the author(s) only and do not necessarily reflect those of the European Union or the European Research Council. Neither the European Union nor the granting authority can be held responsible for them.}

%
\author[1]{Leo P. Mulholland}[type=editor,
                        auid=000,bioid=1,
                        orcid=0009-0003-2668-5589]

\cormark[1]


\ead{lmulholland25@qub.ac.uk}


\credit{Conceptualization of this study, Methodology, Software}

\affiliation[1]{organization={Astrophysics Research Centre,The Queen's University of Belfast},
    addressline={University Road}, 
    city={Belfast},
    postcode={BT71NN}, 
    country={United Kingdom}}
    
\affiliation[2]{organization={Department of Physics, Auburn University},
    addressline={Leach Science Center}, 
    city={Auburn},
    postcode={, AL 36849}, 
    country={United States of America}}
\author[2]{Steven J. Bromley}[style=english]

\author[1]{Connor P. Ballance}[style=english]


\author[1]{Stuart A. Sim}[style=english]
\author[1]{Catherine A. Ramsbottom}[style=english]
\cortext[cor1]{Corresponding author}

\begin{abstract}
The Axelrod approximation is widely used in astrophysical modelling codes to evaluate electron-impact excitation effective collision strengths for forbidden transitions. Approximate methods such as this are a necessity for many heavy elements with open shells where collisional data is either non existent or sparse as the use of more robust methods prove prohibitively expensive. Atomic data for such forbidden transitions are essential for producing full collisional radiative models that do not assume Local-Thermodynamic-Equilibrium (LTE). In this short work we present the re-optimization the simple Axelrod formula for a large number of $R$-matrix data sets, ranging from Fe and Ni to the first r-process peak elements of Sr, Y and Zr, to higher $Z$ systems Te, W, Pt and Au. We show that the approximate treatment of forbidden transitions can be a significant source of inaccuracy in such collisional radiative models. We find a large variance of the optimized coefficients for differing systems and charge states, although some general trends can be seen based on the orbital structure of the ground-state-configurations. These trends could potentially inform better estimates for future calculations for elements where $R$-matrix data is not available.
\end{abstract}


\maketitle

\section{Introduction}
\noindent
The accurate calculation of atomic data for collision processes is computationally expensive and typically has to be carried out on a case-by-case basis for each ion and ionization stage of interest. For this reason, in models where the approximation of Local-Thermodynamic-Equilibrium (LTE) cannot be assumed, the employed codes typically make use of approximate or semi-empirical datasets. The formulae of \citet{Axelrod1980}  and \citet{van1962rate} are often used when calculating the rates of electron-impact-excitation. Examples of works making use of these approximations include \cite{PognanNLTE,pognan2024actinidesignatureslowelectron,shingles2020monte}. \nocite{burke2011r} 
While such formulae are valid for producing usable data for the simulation of astrophysical transients, it has been shown recently by \citet{bromley2023electron} and \citet{mulholland2024_sr_y_ii} that the data these methods produce in their current state are typically inaccurate for many transitions. Generally for forbidden lines the collision strengths are underestimated by several orders of magnitude when compared with results from the $R$-matrix method of \citet{burke2011r}.  Additionally, at low temperatures the $R$-matrix method is more reliable when compared with methods such as distorted-wave  due to the inclusion of resonances in the low energy interaction region. The inclusion of resonances is dependent on the specific atomic system, which increases computational complexity - however their neglection can potentially reduce the calculated rates by a significant margin.  

Recently, a large amount of atomic data has become available for heavy elements of astrophysical interest \cite{dougan2025strontiumiiiiiv,mulholland2024_sr_y_ii,Mulholland24Te,RamsbottomYttrium,smyth2019towards,badnell2014electron,dunleavy2021electron,blondin23,fernandez2019spectroscopic,McCannZr,smyth2018dirac,Dunleavy2022,mccann2024electron,mccann2022atomic,bromley2023electron}, particularly those for kilonovae (KNe), supernovae (SNe) and fusion applications. In this work we compare optically thin emission spectra calculated using the \citet{Axelrod1980}  and \citet{van1962rate} empirical formulae with models based on data produced by sophisticated $R$-matrix calculations. In particular,  the validity of the Axelrod formula for forbidden transitions in the context of modelling KNe is investigated.  Use of the Axelrod and van Regemorter formulae has been discussed previously \cite{bromley2023electron,McCann24,mulholland2024_sr_y_ii}, and some revision has also been put forward by \citet{pognan2024actinidesignatureslowelectron}. In this publication we add to these discussions by presenting an ion-dependent re-optimization of the numerical constant within the Axelrod formula based on the availability of the new $R$-matrix data. The work here serves to illustrate the degree in variation in characteristic collision strengths between different ions, and quantify the extent to which results from full calculations can be captured with such simple formulae. We also discuss how existing data can serve to guide better informed approximation for elements where no $R$-matrix data is available.

In Section \ref{sec:theory}, the theory of collisional-radiative-modelling and electron-impact-excitation is briefly introduced, as well as the approximate formulae of \citet{van1962rate} and \citet{Axelrod1980}. It is also demonstrated, by way of illustrative examples, how the Axelrod formula in its unaltered form can systematically under predict emission from ions. In Section \ref{sec:reopt}, we detail the re-optimization of the coefficients in the Axelrod formula, where we optimize toward newly available calibrated $R$-matrix data. Optically thin emission spectra are constructed using a selection of datasets to show how the Axelrod formula can be amended with an ion-dependent coefficient.  Some basic trends in terms of the shell-behaviour of the ions considered are further analysed. In Section \ref{sec:conc} we conclude with a summary of the main findings of the paper and an outlook.

\section{Theory }\label{sec:theory}
\noindent
Collisional radiative modelling codes solve the rate equations for the level populations  of ions in a plasma,
\begin{equation}
    \frac{\dd N_i}{\dd t} = \sum_{j} C_{ij} N_j,\label{eq:cr}
\end{equation}
where the matrix $C_{ij}$ encapsulates the rates of the atomic processes involved. This includes the electron-impact excitation and de-excitation rates given by
\begin{align}
       q_{i\to j}(T_e) &= \frac{8.63\times10^{-6}}{g_i T_e^{1/2}} \Upsilon_{ij}(T_e) \; e^{-E_{ij}/kT_e} \hspace{1mm}\label{eq:uprates}, \\
    q_{j \to i}(T_e) &= \frac{g_i}{g_j} e^{E_{ij} /kT_e} q_{i\to j}\label{eq:downrates},
\end{align}
where $T_e$ is the electron temperature in Kelvin, $k$ is the Boltzmann constant, $g_{i(j)}$ is the statistical weight of the lower (upper) level and $E_{ij}$ is the energy difference between the levels. The quantities $\Upsilon_{ij}$ represent Maxwellian-averaged-collision strengths. These have been calculated a number of ways, including distorted wave \cite{itikawa1986distorted} and the R-matrix codes \cite{burke2011r}. As discussed in the introduction, in lieu of data produced by these methods, semi-empirical formulae are employed, such as those presented in \cite{van1962rate,Axelrod1980}. The focus of this work is on forbidden lines, for which modelling codes typically employ the formula of \cite{Axelrod1980},
\begin{equation}
    \Upsilon_{ij} = 0.004 g_i g_j \label{eq:axel},
\end{equation}
which was optimized on data from \cite{Garstangetal1978_axelrod_referenced}. This coefficient of 0.004 was specifically produced for data of Fe, and will be shown to result in a systematic underestimation of atomic line emission. The presented formula is independent of temperature, which is discussed in \cite{burgess1992analysis}.

\section{Reoptimization}\label{sec:reopt}
\vspace{2.0mm}
\noindent
We reoptimise the Axelrod approximation for a particular element by fitting the formula,
\begin{equation}
    \Upsilon_{ij} = \alpha g_i g_j, \label{eq:axelgen}
\end{equation}
with lower and upper levels $i$ and $j$ respectively, and corresponding statistical weights $g_i$ and $g_j$. Due to recently published $R$-matrix data for several heavy systems \cite{dougan2025strontiumiiiiiv,mulholland2024_sr_y_ii,Mulholland24Te,RamsbottomYttrium,smyth2019towards,badnell2014electron,dunleavy2021electron,blondin23,fernandez2019spectroscopic,McCannZr,smyth2018dirac,Dunleavy2022,mccann2024electron,mccann2022atomic,bromley2023electron}, it is now possible to reoptimize this coefficient for a large number of systems, retaining Equation \eqref{eq:axelgen} as the formula used for forbidden transitions. In particular, we find the optimal $\alpha$ on an ion-by-ion basis. 

\subsection{Methods}
\vspace{2.0mm}
\noindent 
We define two cost functions for optimising the Axelrod coefficient as,
\begin{align}
    A(\alpha) &= \sum_{k} \sum_{j>i} \big|\Upsilon^{RM}_{ij}(T_k) - \alpha g_i g_j\big|^2 W_{ij}(T_k), \label{eq:my_cost1}\\
    B(\alpha) &= \sum_{k} \sum_{j>i} \big|\log\Upsilon^{RM}_{ij}(T_k) - \log\alpha g_i g_j\big|^2 W_{ij}(T_k)\label{eq:my_cost2},
\end{align}
where the inner sum is taken over all transitions with upper and lower levels $j$ and $i$ respectively and the outer sum is taken over all temperature points. We have also included optimisation with respect to a weight matrix $W_{ij}(T_k)$, where allowed transitions $j$ have  $W_{ij}(T_k) = 0$ identically. This is so that only forbidden transitions are included in the optimization - as allowed transitions are typically calculated in other approximate manners such as \citet{van1962rate}. 

A forbidden transition is sometimes defined in the literature as one with an oscillator strength $f_{i\to j} < 10^{-3}$ \citep{shingles2020monte,PognanNLTE}, although there is a more rigorous definition based on the quantum numbers of the states involved in the transition\footnote{
Technically, allowed transitions are those with a change in angular momentum $\Delta J = 0,\pm1$ ($0 \not\to 0$) and a parity change $\Delta \pi = \pm 1$. A forbidden transition is then one which does not satisfy these selection rules. However, for consistency with the use of the Axelrod formula in the literature, we employ this definition based on the oscillator strength.}. In optimizing against $R$-matrix data, it was found this criterion based on solely the oscillator was too lenient, as some allowed transitions can exhibit large $\Upsilon$ values with small oscillator strengths. For this reason we introduce the additional requirement that $\Upsilon < 100$ on average for a forbidden transition. We therefore identify a forbidden line based on both the condition on the oscillator strength as well as the respective $R$-matrix $\Upsilon$ values.  Additionally, since the primary motivation of this work is the use of Axelrod in late-stage KNe modelling, we restrict the optimization to low temperatures. We thus set $W_{ij}(T_k > 10,000K) = 0$ to roughly encapsulate the temperature range of concern. Both optimizations $A(\alpha)$ and $B(\alpha)$ naturally amount to fitting a straight line through the origin, although we show both to emphasise the heavy dependence of the optimization on the weighting of the included transitions.

\subsection{Results}
\vspace{2.0mm}
\begin{figure*}[ht!]
\centering
    \includegraphics[width=\textwidth]{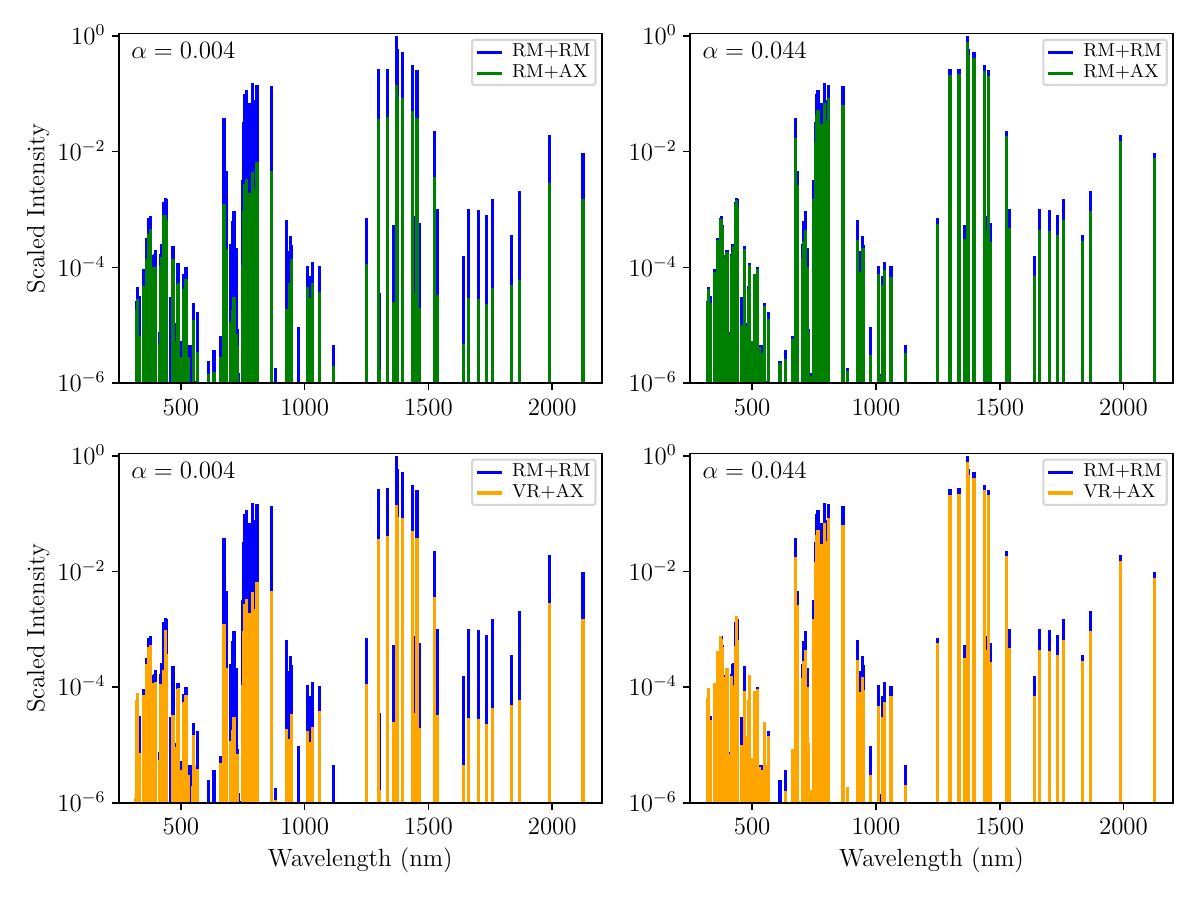}
    \caption{Optically thin emission spectra of Y {\sc ii} at an electron temperature of $2900$ K and electron density $10^6$ cm$^{-3}$. }
    \label{fig:yii_emission_modified}
\end{figure*}

\begin{figure*}[ht!]
\centering
    \includegraphics[width=\textwidth]{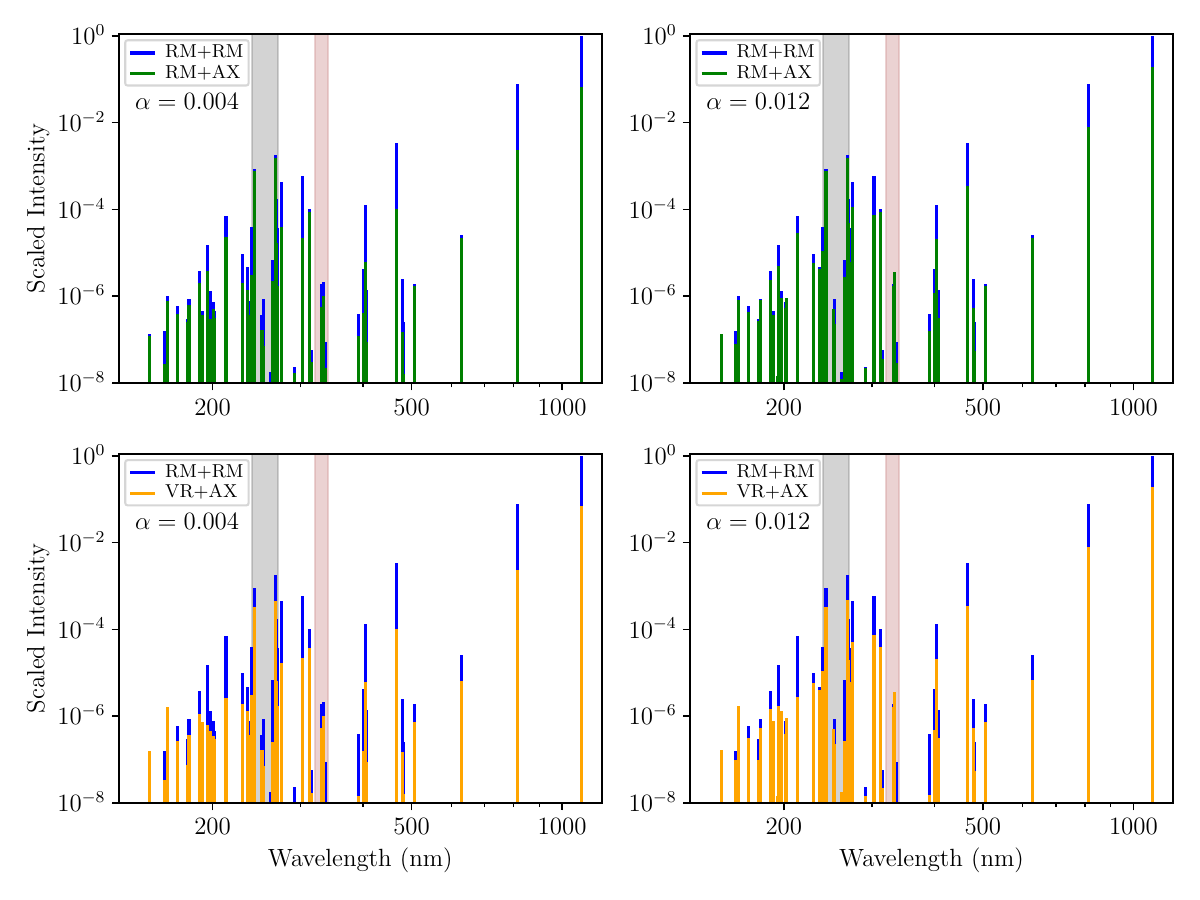}
    \caption{Optically thin emission spectra of Au {\sc i} at an electron temperature of $5800$ K and electron density $10^6$ cm$^{-3}$.}
    \label{fig:aui_emission_modified}
\end{figure*}

The Axelrod approximation was optimised for a number of atomic species for which R-matrix data was readily available. Reoptimized $\alpha$ coefficients are shown in Table \ref{tab:alpha}. The values of $\alpha_A$ are those that minimized the function in Eq. \eqref{eq:my_cost1} and the values of $\alpha_B$ are those that minimized the function in Eq. \eqref{eq:my_cost2}.  Elements are grouped by the highest orbital angular momentum in their ground state configuration. It can be seen that quite generally $\alpha_A$ is larger than $\alpha_B$ for each element, in some cases by an order of magnitude. Given that the Axelrod formula is typically used for the \emph{ad hoc} representation of the general behaviour within a collisional radiative model, the larger values found in $\alpha_A$ will more accurately represent the cooling.   It is seen from Table \ref{tab:alpha} that the $s$-shell elements generally require a factor 10 - 100 times larger than that of the standard Axelrod formula.
Similarly, $p$-shell elements require a factor around 10-20 times higher. By contrast, the $d$-shell elements appear well represented in order of magnitude by the stock formula. This is perhaps expected given the Axelrod formula \cite{Axelrod1980} was originally optimized on the Fe {\sc iii} (3d$^6$) data of \citet{Garstangetal1978_axelrod_referenced}. With expanding atomic data sets, it will be interesting to see the impact of $f$-shell elements when large amounts of atomic data become available for them. Additionally, we have presented the mean values of $\alpha_A$ within the table, for a more universal replacement in models for ions where accurate data is not yet available. Use of these mean values should come with caution, as there remains large spread within the shell groups.

\setlength{\tabcolsep}{3pt}
\begin{table*}[!ht]
    \centering
    \caption{Optimized Axelrod-coefficients for a selection of ion stages with R-matrix $\Upsilon$ data. The data is broken down by shell-angular momentum of the ground state, except for Y {\sc iii}, which is grouped with the s-shell elements due to its effective-single-electron structure. The data is formatted as $a(b) = a \times 10^b$.}
    \label{tab:alpha}
     \begin{tabular}{llcclllccl}
        \hline
        Ion              & Config        & $\alpha_A$& $\alpha_B$& Source& Ion              & Config           & $\alpha_A$& $\alpha_B$& Source\\
        \hline 
        $s$-shell        &               & \textbf{2.4$({-1})$}        &         &        & $d$-shell        & \emph{continued}      &         &         & \setlength{\tabcolsep}{20pt} \\
        Sr {\sc i  }     & 4p$^6$ 5s$^2$ & 5.4$({-2})$ & 2.6$({-2})$ &\cite{dougan2025strontiumiiiiiv} & Ni {\sc ii }     & 3d$^{9 }$         & 2.9$({-3})$   &	$3.7(-4)$ & \cite{dunleavy2021electron} \\  
        Y  {\sc ii }     & 4p$^6$ 5s$^2$ & 4.5$({-2})$ & 3.2$({-2})$ &\cite{mulholland2024_sr_y_ii}    & Ni {\sc iii}     & 3d$({8 })$        & 6.7$({-3})$ &	$2.0({-3})$ & \cite{blondin23} \\ 
        Sr {\sc ii }     & 4p$^6$ 5s     & 4.8$({-1})$ & 2.1$({-1})$ &\cite{mulholland2024_sr_y_ii}    & Ni {\sc iv }     & 3d$({7 })$        & 4.7$({-3})$ &	$1.2({-3})$ & \cite{fernandez2019spectroscopic} \\ 
        Y  {\sc iii}     & 4p$^6$ 4d     & 3.7$({-1})$ & 2.3$({-1})$ &\cite{RamsbottomYttrium}         & Zr {\sc i  }     & 4d$({2 })$ 5s$^2$ & 5.0$({-3})$ &	$2.1({-3})$ & \cite{McCannZr} \\ 
                         &               &         &         &                                         & Zr {\sc ii }     & 4d$^{2 }$ 5s      & 1.7$({-2})$ &	$6.3({-3})$ & \cite{McCannZr} \\ 
        $p$-shell        &               &\textbf{7.7$({-2})$}         &         &                      & Zr {\sc iii}     & 4d$^{2 }$         & 5.0$({-2})$ &	$2.1({-2})$ & \cite{McCannZr} \\ 
        Sr {\sc iii}     & 4p$^6$        & 7.7$(-2)$ &	4.2$({-2})$ &\cite{dougan2025strontiumiiiiiv}  & W  {\sc i  }     & 6s$^{2 }$ 5d$^4$  & 2.8$({-3})$ &	$1.4({-3})$ & \cite{smyth2018dirac} \\ 
        Te {\sc i  }     & 5s$^2$ 5p$^4$ & 6.7$(-2)$ &	1.2$({-2})$ &\cite{Mulholland24Te}             & W  {\sc ii }     & 6s$^{2 }$ 5d$^3$  & 3.7$({-3})$ &	$2.0({-3})$ & \cite{Dunleavy2022} \\ 
        Te {\sc ii }     & 5s$^2$ 5p$^3$ & 7.5$(-2)$ &	5.8$({-2})$ &\cite{Mulholland24Te}             & W  {\sc iii}     & 6s$^{2 }$ 5d$^2$  & 6.5$({-3})$ &	$4.3({-3})$ & \cite{mccann2024electron} \\ 
        Te {\sc iii}     & 5s$^2$ 5p$^2$ & 8.9$(-2)$ &	6.5$({-2})$ &\cite{Mulholland24Te}             & Pt {\sc i  }     & 5d$^{9 }$ 6s      & 5.4$({-3})$ &	$1.5({-3})$ & \cite{bromley2023electron} \\ 
                         &               &         &         &                                         & Pt {\sc ii }     & 5d$^{9 }$         & 4.4$({-3})$ &	$1.7({-3})$ & \cite{bromley2023electron} \\ 
        $d$-shell        &               &\textbf{8.1$({-3})$}         &         &                      & Pt {\sc iii}     & 5d$^{8 }$         & 3.1$({-3})$ &	$1.5({-3})$ & \cite{bromley2023electron} \\ 
        Fe {\sc ii }     & 3d$^{6 }$ 4s  & 2.3$({-3})$ &	5.8$({-4})$ &\cite{smyth2019towards}       & Au {\sc i  }     & 5d$^{10}$ 6s      & 1.2$({-2})$ &	$5.5({-3})$ & \cite{mccann2022atomic} \\ 
        Fe {\sc iii}     & 3d$^{6 }$     & 2.7$({-3})$ &	7.5$({-4})$ &\cite{badnell2014electron}    & Au {\sc ii }     & 5d$^{10}$         & 4.7$({-3})$ &	$5.5({-4})$ & \cite{mccann2022atomic} \\ 
                         &               &         &         &                                         & Au {\sc iii}     & 5d$^{9 }$         & 3.7$({-3})$ &	$1.8({-3})$ & \cite{mccann2022atomic} \\                              \hline
    \end{tabular}
\end{table*}
The emission spectrum as a function of wavelength in nm (scaled relative to the strongest line using only $R$-matrix data) for Y {\sc ii} is shown on Figure \ref{fig:yii_emission_modified}. The left hand panel depicts three optically thin emission spectra using different Y {\sc ii} datasets. Namely, the full R-matrix data for both allowed and forbidden transitions (RM+RM), the R-matrix data for electric dipoles and the baseline Axelrod formula ($\alpha$=0.004) for forbidden lines (RM+AX) and finally the van Regemorter formula for allowed lines and the baseline Axelrod formula for forbidden lines (VR+AX). By contrast, the right hand panel again uses the full R-matrix data (RM+RM), the R-matrix data for electric dipoles and the modified Axelrod formula ($\alpha_A$=0.044) for forbidden lines (RM+AX) and finally the van Regemorter formula for allowed lines and the modified Axelrod coefficient for forbidden lines (VR+AX). It is clear that the emission spectra are more comparable when the modified Axelrod approximation is adopted. It is of note that the agreement increases with only a change to the approximate formula for forbidden transitions. This is an indication that the accuracy of NLTE collisional radiative models is limited by the  treatment of forbidden transitions. This is perhaps not surprising, as forbidden transitions provide efficient thermal excitation/de-excitation pathways at low electron temperatures.

We additionally show on Figure \ref{fig:aui_emission_modified} the scaled emission for Au {\sc i}, employing a similar set of modified data to the above. Au {\sc i} presents a special case of a $d$-shell element whose electronic configurations are constructed from the 5d$^{10}$ $nl$ and 5d$^{9}$ $nl n'l'$ series. The former set behaves like a single electron system. There is therefore a limited number of forbidden transitions that are typically stronger than those of other systems. For this reason the optimized Axelrod coefficient is considerably higher than the surrounding $d$-shell elements presented in Table \ref{tab:alpha}. Figure \ref{fig:aui_emission_modified} therefore demonstrates a lack of consistency in terms agreement in the emission estimations because the Axelrod formula assumes that all the forbidden lines behave the same, which is evidently not the case for this ion.

It can be seen that certain lines are well represented with the R-matrix + Axelrod datasets (RM+AX), regardless of the Axelrod coefficient used. It was found these represent dipole allowed lines whose upper level population is dominated by the corresponding excitation rate, which are represented by the R-matrix data in that formalism. It is noteworthy that such lines are typically poorly represented by the van Regemorter + Axelrod data (VR+AX), showing the limiting factor for these lines at this point in temperature and density space are actually the effective collision strengths of allowed transitions. Two examples can be seen at $\sim 250$nm, which are highlighted in grey on Figure \ref{fig:aui_emission_modified}. Notably both mixes of data show little benefit from the improved Axelrod approximation. In fact, those lines that were truly dominated by the R-matrix dipole data are in some cases over-estimated by the improved Axelrod approximation (see for example the lines between 300 and 400nm highlighted in red). 

By contrast (and is the case for the majority of the lines shown) there is significant underestimation of the emission in the case of the base Axelrod approximation, where agreement is seen between the data sets employing this with either the R-matrix or van Regemorter allowed rates. Moving to the re-optimized Axelrod approximation introduces a certain improvement in the estimation of the emission, in many cases by an order of magnitude. The primary result is the improvement of the evidently strongest lines which are most likely to contribute to a real astrophysical spectrum.

In principle we could optimise only on the strongest lines, although the optimization would depend on the plasma conditions chosen. For a low density plasma, the coronal approximation applies and the optimization would reduce to fitting Eq. \eqref{eq:axelgen} to the subset of effective collision strengths $\Upsilon_{1i}$. For a high density plasma, LTE applies and in principle the populations and thus the emission should be independent of the rates. Given that the particular positions in density space of the coronal, collisional-radiative and thermal equilibrium regimes are vastly different for different ion stages, perhaps a general optimization over all the available rates as performed here remains the best solution as opposed to the choice of particular lines. Such an effort may be feasible when a greater breadth of $R$-matrix data spanning the full range of $s-$, $p-$, $d-$, and $f-$shell elements relevant to KNe becomes available.

\section{Conclusions and Outlook} \label{sec:conc}

To summarise, the Axelrod approximation for effective collision strengths has been reviewed and revisited. The Axelrod coefficient has been calculated for a variety of ion stages with available R-matrix data. The elements studied cover the 1st (Sr-Zr), 2nd (Te) and 3rd (W,Pt,Au) peaks of the rapid $(r)$ nucleosynthesis processes. 
The values presented here could in principle be used to generate collision strengths for collisions between high-lying levels not included in structure calculations, or aid in the calculation of similarly structured elements by using approximate coefficients reflective of those presented in Table \ref{tab:alpha}. Elements with $s$-shell ground states tend to have transitions 10 - 1000 times stronger than the base Axelrod approximation. $p$-shell ground states tend to be 10-20 times stronger. $d$-shell elements tend to have forbidden transitions of the rough order of magnitude of the original Axelrod approximation. Despite this, the relatively large variance in $\alpha$ indicates that the Axelrod approximation should be reserved for where \emph{no} data is available, and should be replaced with $R$-matrix data as it comes available. Nonetheless, it is intended that this work serves as an intermediate step in the improvement of atomic data for astrophysical models, where the $\alpha$ values (or perhaps their mean values) presented here can be used to provide more information for elements with no data available and the Axelrod formula can be scaled based on the electronic configurations of the species involved.

\section*{CRediT authorship contribution statement}
\textbf{Leo P. Mulholland:} Writing – review \& editing, Writing – original draft,
Visualization, Validation, Methodology, Investigation, Conceptualization, Formal analysis.

\textbf{Steven J. Bromley:} Writing – review \& editing, Writing – original draft,
Visualization, Validation, Methodology, Investigation, Conceptualization, Formal analysis.

\textbf{Connor P. Ballance:} Writing – review \& editing, Writing – original draft, Validation, Supervision.

\textbf{Stuart A. Sim:} Writing – review \& editing, Writing – original draft, Validation, Methodology, Visualization, Supervision.

\textbf{Catherine A. Ramsbottom:} Writing – review \& editing, Writing – original draft, Validation, Methodology, Visualization, Supervision.

\section*{Declaration of competing interest}
\noindent The authors declare that they have no known competing financial interests or personal relationships that could have appeared to influence the work reported in this paper.

\section*{Data Availability}
\noindent The relevant atomic data can be found in the cited references.

\section*{Acknowledgements}\label{sec:acknowledgements}
\noindent We thank our colleagues at Queen's University Belfast and Auburn University for helpful discussion. LPM, CPB, SAS and CAR acknowledge funding from the European Union (ERC, HEAVYMETAL, 101071865). Views and opinions expressed are however those of the author(s) only and do not necessarily reflect those of the European Union or the European Research Council. Neither the European Union nor the granting authority can be held responsible for them. SJB acknowledges NASA APRA for funding part of this work under grant number is 23-APRA23-0086.



\bibliographystyle{unsrtnat}

\bibliography{bibliography}

\end{document}